# Subcycle Quantum Electrodynamics

C. Riek[1], P. Sulzer[1], M. Seeger[1], A. S. Moskalenko[1], G. Burkard[1], D. V. Seletskiy[1], and A. Leitenstorfer[1]

[1] Department of Physics and Center for Applied Photonics, University of Konstanz, D-78457 Konstanz, Germany

**Besides their stunning physical properties which are unmatched in a classical world, squeezed states[1-4] of electromagnetic radiation bear advanced application potentials in quantum information systems[5] and precision metrology[6], including gravitational wave detectors with unprecedented sensitivity[7]. Since the first experiments on such nonclassical light[8,9], quantum analysis has been based on homodyning techniques and photon correlation measurements[10,11]. These methods require a well-defined carrier frequency and photons contained in a quantum state need to be absorbed or amplified. They currently function in the visible to near-infrared and microwave[12] spectral ranges. Quantum nondemolition experiments may be performed[13,14] at the expense of excess fluctuations in another quadrature. Here we generate mid-infrared time-locked patterns of squeezed vacuum noise. After propagation through free space, the quantum fluctuations of the electric field are studied in the time domain by electro-optic sampling with few-femtosecond laser pulses[15,16]. We directly compare the local noise amplitude to the level of bare vacuum fluctuations. This nonlinear approach operates off resonance without absorption or amplification of the field that is investigated. Subcycle intervals with noise level significantly below the pure quantum vacuum are found. Enhanced fluctuations in adjacent time segments manifest generation of highly correlated quantum radiation as a consequence of the uncertainty principle. Together with efforts in the far infrared[17,18], this work opens a window to the elementary quantum dynamics of light and matter in an energy range at the boundary between vacuum and thermal background conditions.**

Coherent states represent the closest counterpart to a classical electromagnetic wave that exists in quantum electrodynamics. The quantum noise amplitudes of their electric and magnetic fields coincide precisely with those of the vacuum state[19]. Recently, we have succeeded to directly detect the bare vacuum fluctuations of the mid-infrared electric field with highly sensitive electro-optic sampling based on ultrashort laser pulses[15,16]. One key aspect of this technique is that it operates out of a time-domain perspective. Therefore, it should provide a resolution substantially below the duration of an oscillation period of any quantum field under study. Naturally, it is tempting to think about an experiment that synchronously couples a nonclassical state of light into the space-time volume which is probed, thus providing a quantum noise amplitude that deviates from pure vacuum fluctuations. Especially, it would be an attractive manifestation of quantum physics if *less noise* as compared to the quantum vacuum could be localized in time and space. In conventional homodyning studies, the carrier wave of a local oscillator needs to be phase-locked to a quantum state[11,16]. Instead, we have to prepare a squeezed electromagnetic transient with a noise pattern that is synchronized with the intensity envelope of an ultrashort probe pulse. This tightly focussed few-femtosecond optical wave packet then defines a subcycle space-time segment in which the quantum statistics of a mid-infrared nonclassical signal is sampled.

Our scheme to implement such an experiment is sketched in Fig. 1(a). We send an intense near-infrared pump pulse (red-yellow envelope) with duration of 12 fs and centre frequency of 200 THz into a thin generation crystal (GX). In a first step, a carrier-envelope phase-locked electric field transient[20] is generated by optical rectification (red line). Once built up, it starts to locally phase shift the co-propagating multi-terahertz vacuum fluctuations (green shaded band) by means of the electro-optic effect in the GX which establishes a change in refractive index $\Delta n(t)$ proportional to the mid-infrared electric field amplitude $E_{THz}(t)$. In a simplified picture, the resulting local anomalies in the speed of light might induce depletion of vacuum amplitude at certain space-time regions (blue shaded sections), piling it up in others (stained in red). A high efficiency for this two-step mechanism to squeeze the mid-infrared vacuum is ensured by the large second-order nonlinearity of the 16-μm-thick exfoliated piece of GaSe we employ as GX[20]. Tight focussing of the pump to a paraxial spot radius $w_{pump}$ of 3.6 μm also defines the transverse spatial mode for the nonclassical electric field pattern. After the GX, the squeezed vacuum is collimated and residual pump is removed by a 70-μm-thick GaSb filter inserted under Brewster's angle. A mode-matched 5.8 fs probe pulse (blue envelope) is then superimposed onto the multi-terahertz field and focussed to $w_{probe}$ = 3.6 μm in a AgGaS$_2$ detector crystal (DX) of 24 μm thickness[15]. It samples the electric field in the co-propagating space-time volume via the electro-optic effect[15,20] and as a function of time delay $t_D$. We gain two different types of information: the coherent ("classical") electric field amplitude $E_{THz}(t_D)$ of the squeezing mid-infrared transient is recorded in the conventional way[20]. In addition, the quantum distribution of the multi-terahertz electric field is accessed via statistical readout[15]. Especially, our technique allows us to directly reference the local noise level $\Delta E_{rms}$ in a squeezed transient (blue and red distributions in Fig. 1(b)) to the fluctuations $\Delta E_{vac}$ obtained under bare vacuum input[21] (green distribution in Fig. 1(b)). Relative differential noise (RDN) patterns mirroring $\Delta E_{rms}$ are then recorded as a function of delay time $t_D$ (see Fig. 1(c)). Note that only 4% of the total fluctuation amplitude in our setup result from a bare multi-terahertz vacuum input while the rest is due to the noise-equivalent field of the detector $\Delta E_{SN}$ caused by the quantized flux of near-infrared probe photons[15,21]. Therefore, a RDN of -0.04 would correspond to a complete removal of the vacuum fluctuations in the space-time segment sampled in the DX.

The coherent field transients $E_{THz}(t_D)$ generated by optical rectification of a near-infrared 12 fs pump with pulse energy of 3.5 nJ are depicted in Fig. 2(a). Two waveforms with precisely inverted amplitudes result from rotation of the pump polarization by 90° around the optical axis (black and grey lines). Broadband amplitude spectra (inset) feature an average frequency of 44 THz, corresponding to a free-space wavelength of 6.8 μm and photon energy of 180 meV. Fig. 2(b) shows the RDN amplitudes recorded simultaneously. Dark (light) blue areas denote delay times with negative values induced by the black (grey) transient in Fig. 2(a), indicating a clear squeezing of the local electric field fluctuations $\Delta E_{rms}$ below the level of the bare quantum vacuum. Time segments carrying excess noise with respect to the vacuum ground state are filled by dark or light red colour, respectively. Salient features in the noise patterns of Fig. 2(b) are evident: (i) There exists a clear asymmetry with positive excess noise surpassing the absolute values of vacuum squeezing, especially in the region close to the centre of the transients where the amplitudes are maximum. (ii) The noise maxima in Fig. 2(b) coincide with the maximally negative slopes of the coherent field amplitudes in Fig. 2(a) while optimum squeezing of $\Delta E_{rms}(t_D)$ is obtained close to the positions with a maximally positive increase of $E_{THz}(t_D)$ with time (see vertical dashed lines). (iii) Due to this inherent polar asymmetry in $\Delta E_{rms}$, the shift of carrier-envelope phase between the black and grey transients in Fig. 2(a) results in distinctly different quantum noise patterns that are not mirror images of each other.

We will now investigate the physical origin of these findings. To this end, we first vary the pulse energy in the near-infrared pump which is proportional to the electric fields $E_{THz}$ and record the resulting *RDN* amplitudes (Fig. 3). At low pump energies of 0.8 and 1.5 nJ, the noise patterns are still fairly symmetric with respect to positive and negative extrema. The asymmetry towards positive excess noise shows up clearly at 2.5 nJ and becomes distinct at 3.5 nJ. The origin of these observations is qualitatively understood in terms of the following expression for $\Delta E_{rms}(t)$ at the exit surface of a GX[21].

$$\Delta E_{rms}(t) = e^{f(t)} \Delta E_{vac} \quad \text{where} \quad f(t) = \frac{dl}{nc} \frac{\partial}{\partial t} E_{THz}(t) \qquad (1)$$

denotes the squeezing factor in the time domain. We adopt plane waves and negligible pump depletion in a medium of second-order nonlinear coefficient $d$ and thickness $l$. The bare vacuum amplitude $\Delta E_{vac}$ is assumed as input, adequate to the quantum properties of the coherent pump. $c$ denotes the speed of light in vacuum and a constant refractive index $n$ is well justified because of the minor dispersion of GaSe in the mid infrared. It is evident from Eq. (1) that the extrema in $\Delta E_{rms}(t)$ are expected at the positions of maximum slope of $E_{THz}(t)$, as confirmed experimentally in Fig. 2. With increasing amplitude of $E_{THz}(t)$, a nonlinear relationship between squeezing and excess noise results with respect to $\Delta E_{vac}$ because of the exponential character, tentatively explaining the build-up of the asymmetry in Fig. 3. We now select two points in time $t_{max}$ and $t_{min}$ with opposite slope of $E_{THz}(t)$, i.e. $f(t_{max}) = |f(t_{max})| = -f(t_{min})$. With Eq. (1) and the quantitative expression for the vacuum amplitude[16],

$$\Delta E_{rms}(t_{min}) \cdot \Delta E_{rms}(t_{max}) = (\Delta E_{vac})^2 = \frac{\hbar}{\varepsilon_0 \Delta x \Delta y \Delta z \Delta t} \quad (2)$$

results where a four-dimensional space-time segment is defined by the transverse modal cross section $\Delta x \Delta y = w_{probe}^2 \pi$ and the effective spatio-temporal length $\Delta z \Delta t$ set by the intensity envelope of the probe[15]. $\hbar$ is the reduced Planck constant and $\varepsilon_0$ the permittivity of free space. In order to experimentally verify Eq. (2), two delay times with extremal time derivatives of $E_{THz}(t)$ are sampled. We plot the measured values for $RDN(t_{max})$ and $RDN(t_{min})$ versus near-infrared pump pulse energy (red and blue circles in Fig. 4, respectively). The green graphs represent a least-square fit to those data based solely on Eq. (2). A saturation behaviour is found for squeezing and a superlinear increase for anti-squeezing, in good agreement between experiment and theory. The free parameter determining the asymmetry between the green graphs in Fig. 4 may be exploited to calibrate the amount of squeezing achieved in the experiment (green scale on the right). A value for $1 - \exp(f(t_{min}))$ close to 50% is obtained at a pump energy of 3.5 nJ, corresponding to a decrease of $RDN$ amplitude in the electro-optic signal by approximately $10^{-2}$ (left abscissa in Fig. 4). Note that the asymmetry originates from the maximum squeezing of the mid-infrared quantum field that is achieved inside the GX. Spurious reflections at the uncoated surfaces of GX and DX as well as imperfect segment matching to the spatio-temporal probe wave packet will contaminate the nonclassical state with bare vacuum noise. Therefore, the 50% of local squeezing inside the GX inferred by the analysis above and a resulting $RDN$ of -0.01 (see Fig. 4) are soundly compatible with the maximal noise change of -0.04 that would result under complete suppression of mid-infrared vacuum fluctuations in the DX.

A detailed consideration of the physical character and origin of the squeezed wave packets we sample is rewarding. From Fig. 2 and the discussion around Eq. (1), it is evident that the quantum noise patterns $\Delta E_{rms}(t_D)$ exhibit approximately the same temporal periodicity as the field transients $E_{THz}(t_D)$. Consequently, the total quantum state corresponds to neither amplitude nor phase squeezing of the coherent transients $E_{THz}(t_D)$. To detect nonclassical behaviour with a conventional method such as balanced homodyning, a coherent field centred at half the carrier frequency of $E_{THz}(t_D)$, i.e. around 22 THz, would have to be employed as local oscillator. This situation is analogous to established squeezing experiments based on a second-order nonlinearity and spontaneous parametric fluorescence[9]. Our noise patterns therefore correspond to an ultrabroadband generation of correlated photon pairs[22,23] by $E_{THz}(t_D)$, with total energies distributed symmetrically around the carrier frequency of 44 THz[16]. Owing to the quadratic dependence of the intensity on electric field amplitude, already the symmetric deviations from the vacuum level detected under low pump conditions in Figs. 3 and 4 equal generation of finite energy in the form of highly correlated photon pairs, in agreement with the fact that there exists only one unique ground state[19]. The asymmetry which shows up at higher pump fluences in accord with the exponent in Eq. (1) marks the transition into a regime with higher-order correlations. On subcycle scales, these photons remain indistinguishable and their correlated behaviour at beam splitting elements[24] results in conservation of the ratio between squeezing and excess noise, despite degradation by the partially reflecting facets of GX and DX. Traditionally, squeezing has been discussed in the frequency domain as being due to amplification and de-amplification of specific field quadratures[25]. But the Pockels effect motivated already in the introduction links the term $d \cdot \partial E_{THz}/\partial t$ in Eq. (1) to a modulation of the refractive index $\partial n/\partial t$. Therefore, redistribution of vacuum fluctuations following local advancement and slowdown of the speed of light[26] represents an attractive alternative to illustrate generation of nonclassical radiation in our subcycle time-resolved situation. The time-domain manifestation of Heisenberg's uncertainty principle in Eq. (2) leads to an imbalance between the excess noise related to acceleration of the co-propagating reference frame and squeezing of the quantum amplitude originating from local deceleration[21].

In conclusion, a time-domain perspective on quantum electrodynamics works with subcycle resolution and direct referencing of electric field fluctuations to the quantum vacuum. The high peak intensities provided by few-femtosecond laser pulses of minute energy content enable a compact quantum technology based on broadband nonlinearities without immediate need for enhancement cavities, waveguides or cryogenic cooling. Many stimulating and fundamental questions arise concerning a generalized understanding of quadratures being linked to local accelerations of the moving reference frame or the benefits and limits of the inherently non-destructive character of the technique. Future extensions are to be explored aiming for example at a full quantum tomography[27] on subcycle scales. Filling the gap of quantum approaches in the mid-infrared or multi-terahertz range provides interesting perspectives. Attractive applications including access to new quantum states produced by subcycle perturbation of ultrastrongly coupled light-matter systems[28] or quantum spectroscopy[29] and manipulation of collective degrees of freedom in condensed matter are all inherent to this regime.

**Acknowledgements** The authors wish to thank W. Belzig and D. Brida for stimulating discussions. Support by ERC Advanced Grant 290876 "UltraPhase", by DFG via SFB767 and by NSF via a Postdoc Fellowship for D.V.S. (Award No. 1160764) is gratefully acknowledged.

**Figure Captions**

**Figure 1** *Scheme for time-locked generation and detection of quantum transients by electro-optic sampling.* **a** A 12-fs near-infrared pump pulse (red-yellow envelope) and multi-terahertz vacuum fluctuations (green band) co-propagate into a generation crystal (GX) with second-order nonlinearity. A coherent mid-infrared transient results (red line) which squeezes the quantum vacuum (red and blue sections). A 6-fs probe pulse is superimposed (blue envelope) to sample the electric field amplitude as a function of delay time $t_D$ in an electro-optic detector crystal (DX). **b** Sketch of probability distributions of the electric field at $t_D$ with vanishing coherent amplitude for bare vacuum (green), squeezing (blue) or anti-squeezing (red). **c** Illustration of a trace of relative differential noise (*RDN*) of the quantum field as sampled versus delay time $t_D$.

**Figure 2** *Relative differential noise patterns and dependence on carrier-envelope phase of generating coherent field*. **a** Complementary electric field amplitude of classical mid-infrared transients sampled in the electro-optic detector, as obtained with two different settings of near-infrared pump polarization on the GaSe emitter crystal (black and grey lines). The inset shows the amplitude spectrum. **b** Relative differential noise traces, as recorded together with the classical signals in (a). Sections with less noise as compared to the bare vacuum are depicted in blue and excess noise in red. The strongly (lightly) coloured squeezing pattern with black (grey) envelope corresponds to the black (grey) transient in (a). Vertical dashed lines are to guide the eye towards the phase relationship between (a) and (b).

**Figure 3** *Development of relative differential noise patterns for different near-infrared pump pulse energies $E_{pump}$*. Blue-filled sections indicate squeezing with respect to the bare vacuum electric field amplitude while red areas denote anti-squeezing. An asymmetry between negative and positive values builds up with increasing $E_{pump}$.

**Figure 4** *Build-up of asymmetry between squeezing and anti-squeezing as a consequence of the uncertainty principle.* Extremal values of relative differential noise at a maximally squeezed temporal position (blue dots) and for adjacent excess fluctuations (red dots) are plotted versus near-infrared pump pulse energy. The green line represents a least-square fit to the data based on Eq. (2), yielding the relative squeezing values (right abscissa) which refer to the quantum state inside the generation crystal.

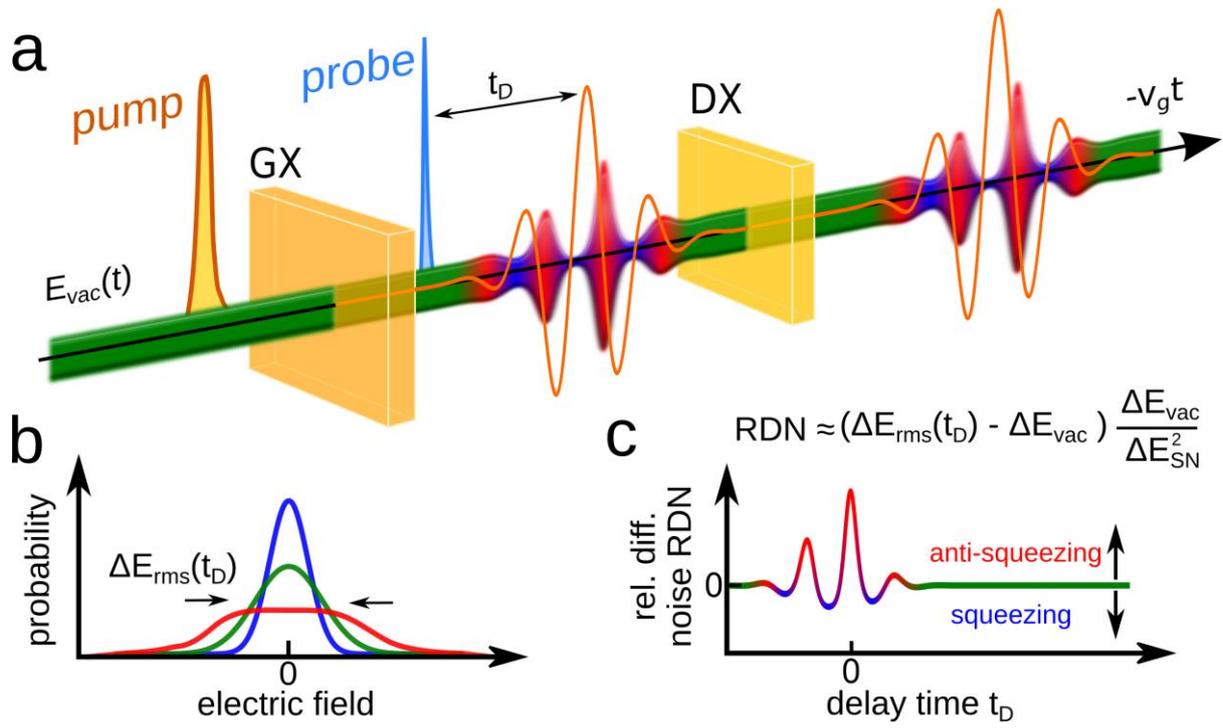

C. Riek *et al.*, Figure 1

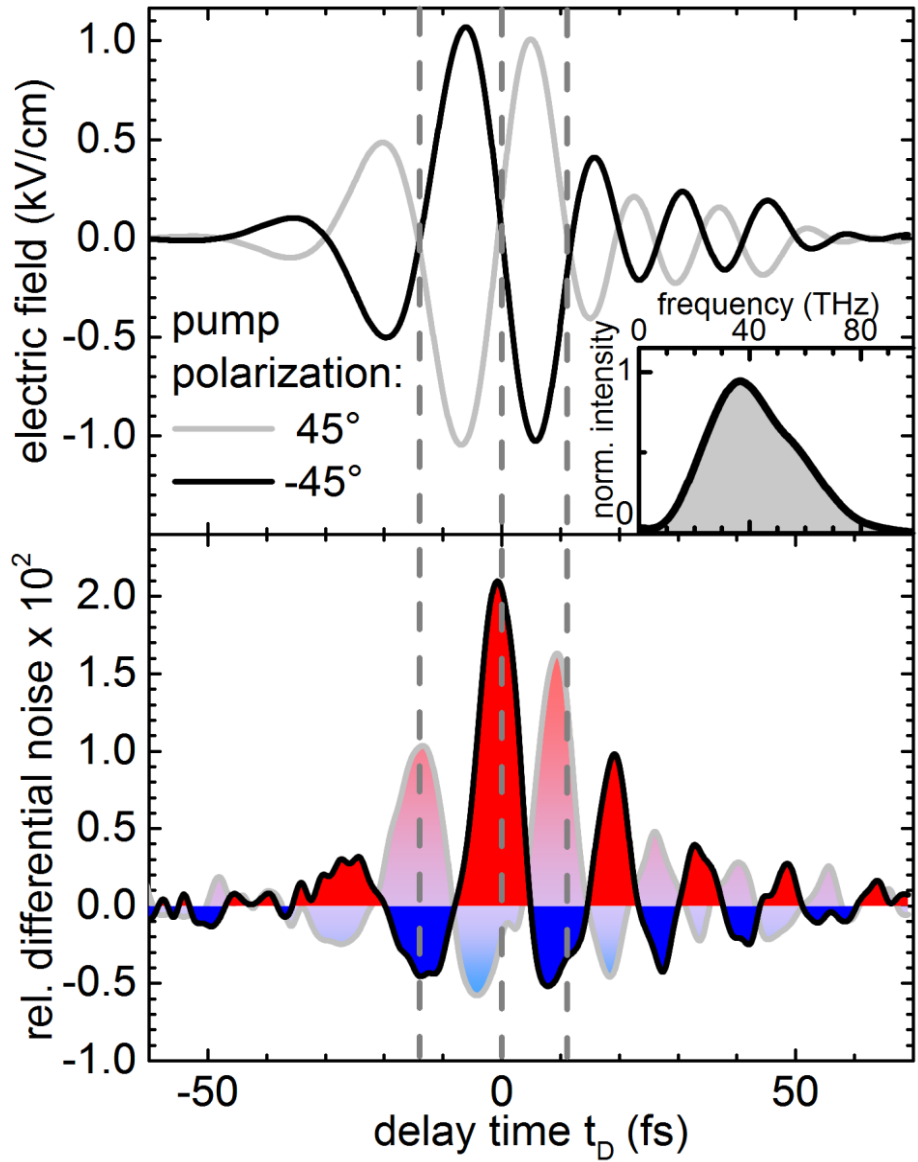

C. Riek *et al.*, Figure 2

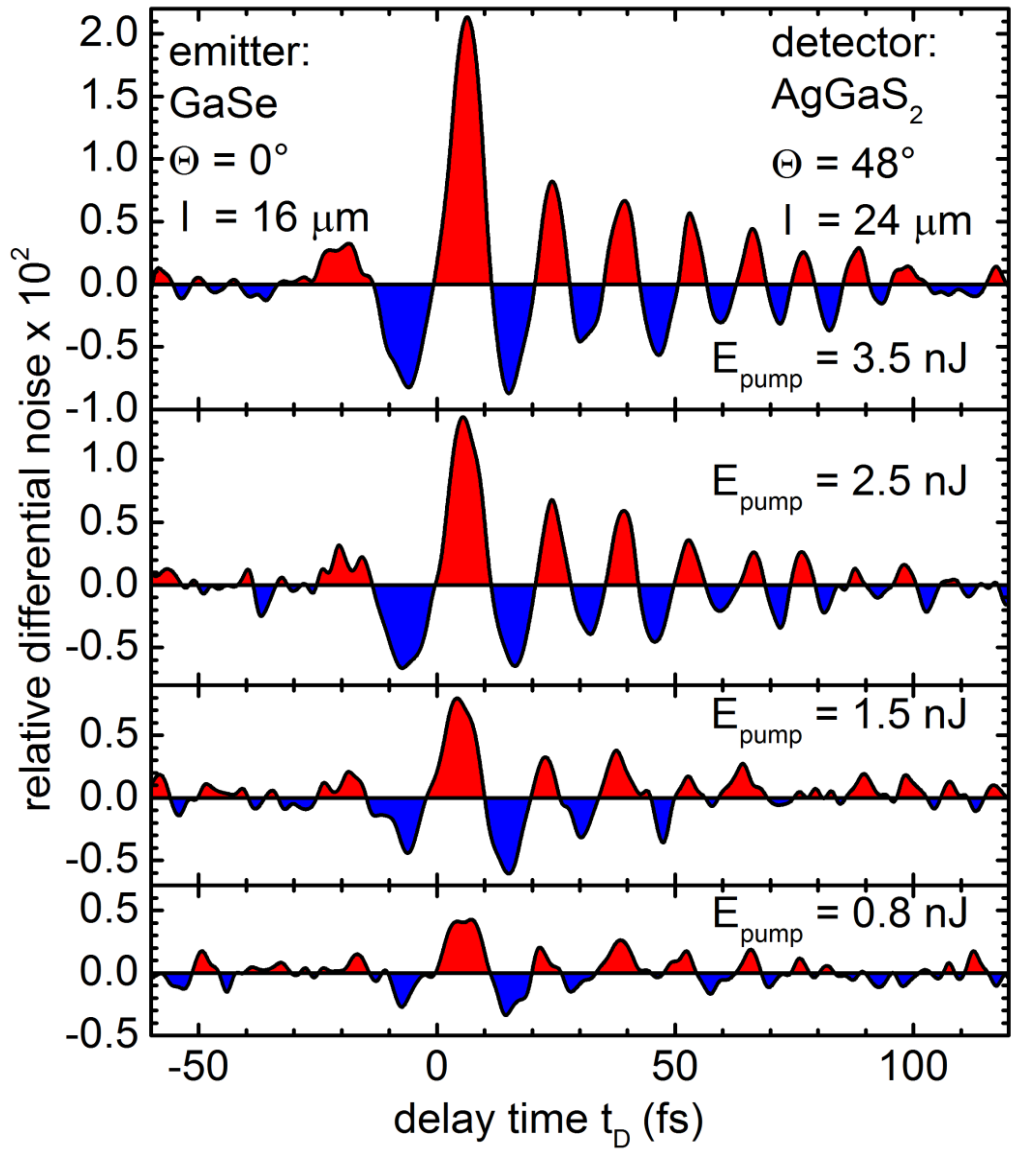

C. Riek *et al.*, Figure 3

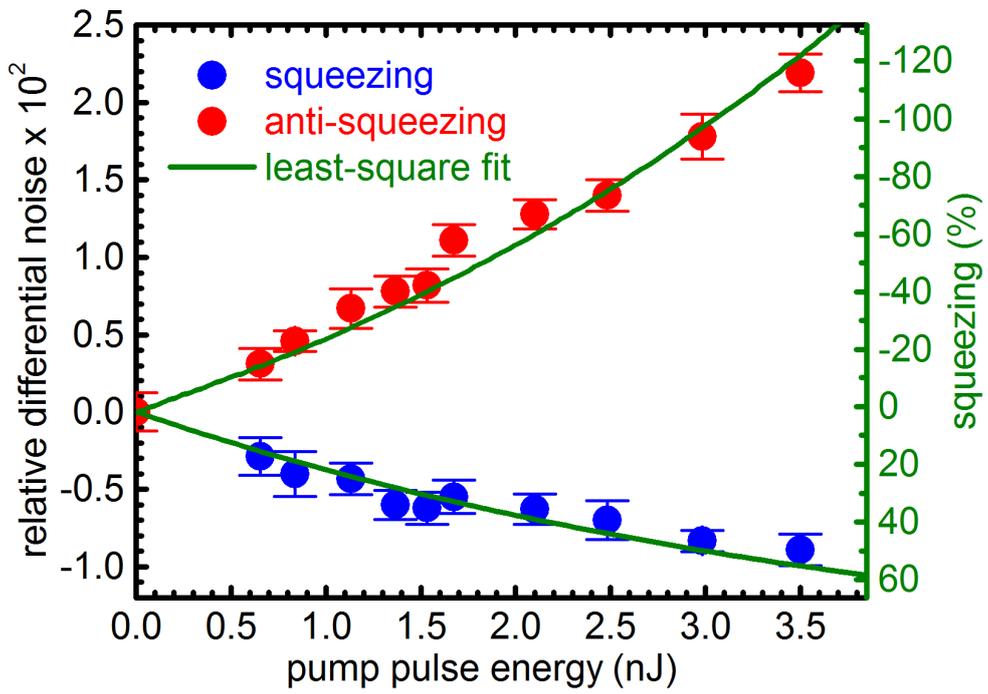

C. Riek *et al.*, Figure 4

**Methods**

**Readout of classical field amplitude and quantum noise in electro-optic sampling:**

Due to the electro-optic (Pockels) effect, the electric field amplitude $E_{THz}$ of an electromagnetic wave propagating in a detector crystal (DX) with second-order nonlinearity leads to a quasi-instantaneous change in refractive index $\Delta n \sim E_{THz}$. The induced birefringence causes a modified polarization state of a co-propagating probe pulse which may be analysed by ellipsometry. In our geometry, we sample only one linear polarization component which is perpendicular to the electric field of the near-infrared probe. The information about the temporal structure of $E_{THz}$ then results from repeated measurements sampling at different time delays $t_D$ between the ultrashort probe and the electric field transient[15,30]. The frequency bandwidth of this method is ultimately limited by the intensity envelope or "pulse duration" of the probe which needs to be close to or shorter than a half-cycle period of $E_{THz}(t_D)$. The noise level of such measurements is determined by the shot noise due to the quantized flux of high-frequency probe photons[16]. We work with a repetition rate of our probe pulses of 40 MHz. Every second pulse is removed from the pump pulse train, resulting in a modulation frequency of 20 MHz where the coherent field amplitude $E_{THz}(t_D)$ is decoded by lock-in detection.

For the quantum noise measurements, the probe may be seen as divided into two sub-pulse-trains with one of them sampling the potentially nonclassical state coming from a synchronized emitter and the other one measuring the bare vacuum noise as a reference. In our setup, we detect both signals simultaneously by taking lock-in measurements in two orthogonal channels locked to a 10 MHz reference input derived from the 20 MHz pump by frequency division. We then compute the root-mean-square noise levels of both sub-readouts and subtract them. The relative differential noise (*RDN*) that is recorded corresponds to the following physical quantity:

$$RDN(t_D) = \frac{\sqrt{\Delta E_{SN}^2 + \Delta E_{rms}^2(t_D)} - \sqrt{\Delta E_{SN}^2 + \Delta E_{vac}^2}}{\sqrt{\Delta E_{SN}^2 + \Delta E_{vac}^2}} \approx \left(\Delta E_{rms}(t_D) - \Delta E_{vac}\right)\frac{\Delta E_{vac}}{\Delta E_{SN}^2}. \quad (3)$$

Here, $\Delta E_{rms}(t_D)$ is the standard deviation of the electric field sampled at a delay time $t_D$, $\Delta E_{vac}$ the fluctuation amplitude of bare vacuum and $\Delta E_{SN}$ the shot-noise equivalent field[15,16] of the electro-optic detection. We compute $\Delta E_{vac} = 24$ V/cm and $\Delta E_{SN} = 81$ V/cm based on Ref. 15 and the slightly modified detection parameters of our present setup. Statistical summation of both values and normalization to the bare shot noise results in a 4% contribution of bare mid-infrared vacuum noise to the fluctuations measured with the electro-optic detection in our present configuration when no squeezed field is present. The linearized approximation on the right part of Eq. (3) enters Fig. 1(c) and holds for the limit of moderate deviations of $\Delta E_{rms}(t_D)$ from $\Delta E_{vac}$. The analysis in Fig. 4 is carried out taking into account the precise relationship between $RDN$ and $\Delta E_{rms}(t_D)$ from the center part of Eq. (3).

Highly synchronous pump and probe pulse trains with minimum amplitude fluctuations are provided by a compact femtosecond Er:fiber laser system[31] which is based entirely on telecom components. By working at the highest possible lock-in frequencies for both the coherent field and $RDN$ readouts, we ensure minimum timing jitter of 1 attosecond between pump and probe[32] as well as quantum-limited amplitude fluctuations. Ultimately, these facts allow us to operate in a regime where any technical noise of the setup is negligible.

Collimation of the classical and quantum fields from the GX is carried out with a gold-coated off-axis parabolic mirror of focal length $f = 15$ mm, the probe pulse is coupled in under s-polarized reflection on a polished Si wafer of 500 μm thickness which is inserted under Brewster's angle for the mid-infrared and an off-axis paraboloid with $f = 15$ mm serves to focus both signal and probe into the DX.

**Theoretical considerations leading to Equation (1):**

The generation of the quantum electric field patterns in our experiment may be understood as a series of two subsequent nonlinear processes of second order. First, a few-femtosecond pump pulse in the near infrared produces an ultrashort and coherent electric field transient $E_{THz}(t)$ at multi-terahertz frequencies $\Omega$ by optical rectification in a generation crystal (GX). This step corresponds to a difference frequency mixing process within the broadband spectrum of the pump, resulting in an identical carrier-envelope phase for multi-terahertz transients in the entire pulse train produced by the mode-locked laser system[20].

In a second step, $E_{THz}(t)$ starts driving the second-order nonlinearity in the GX. We adopt propagation of plane waves in the nonlinear element along the $z$-axis from $-l/2$ to $l/2$ as well as an appropriate mutual orientation of pump field polarization and GX. The one-dimensional picture is well justified in our geometry because the thicknesses of both GX and DX are smaller than the Rayleigh range of the mid-infrared radiation that is generated. Together with the high-NA off-axis parabolic mirrors, this fact ensures proper matching to a single transverse mode. The nonlinear coefficient $d$ is proportional to the second-order nonlinearity $\chi^{(2)}$ of the emitter material[33] and $n$ is the linear refractive index. All susceptibilities may be assumed as dispersionless when the mid-infrared frequencies $\Omega$ are far from the electronic and optical phonon resonances of the medium. In the vacuum picture[34], the total mid-infrared quantum field may be written as $\hat{E}_{THz} = E_{THz} + \delta\hat{E}_{THz}$, with the classical coherent amplitude $E_{THz} = \langle\hat{E}_{THz}\rangle$ and a pure quantum correction $\delta\hat{E}_{THz}$. Locally, $\hat{E}_{THz}$ induces the second-order nonlinear polarization $\hat{P}^{(2)} = -\varepsilon_0 d \cdot \hat{E}_{THz}\hat{E}_{THz}$, acting as a source in the wave equation. We restrict ourselves to small pump depletion by omitting the correction to the classical part and neglect the second-order terms in $\delta\hat{E}_{THz}$. The slowly varying amplitude approximation[33] then leads to

$$\frac{\partial}{\partial z}\delta\hat{E}_{THz}(z,\Omega) = -i\frac{\Omega d}{nc}\int_{-\infty}^{\infty}d\Omega' E^*_{THz}(z,\Omega'-\Omega)\delta\hat{E}_{THz}(z,\Omega'). \quad (4)$$

Transforming back into the time domain and using a modified reference frame with $t' = t - zn/c$, $z' = z$, $\delta\hat{E}'_{THz}(z',t') = \delta\hat{E}_{THz}(z,t)$ and $E'_{THz}(z',t') = E_{THz}(z,t)$, we obtain

$$\frac{\partial}{\partial z'}\delta\hat{E}'_{THz}(z',t') = \frac{d}{nc}\left[\frac{\partial}{\partial t'}E'_{THz}(z',t')\delta\hat{E}'_{THz}(z',t') + E'_{THz}(z',t')\frac{\partial}{\partial t'}\delta\hat{E}'_{THz}(z',t')\right]. \quad (5)$$

As long as deviations of the quantum field from the level of bare vacuum remain moderate, the temporal derivative of $\delta\hat{E}'_{THz}(z',t')$ is negligible and we may omit the second term in the brackets on the right-hand side of Eq. (5). The same term vanishes even for large squeezing when $E'_{THz}(z',t')$ is sufficiently small. In both cases, an analytical solution of the partial differential equation is straightforward by integrating over $z'$. Returning to the original reference frame, the field at the exit surface of the GX, $\delta\hat{E}_{THz,\text{out}}(t) \equiv \delta\hat{E}_{THz}(z=l/2,t)$, may be expressed as

$$\delta\hat{E}_{THz,\text{out}}(t) = e^{f(t)}\delta\hat{E}_{THz,\text{in}}(t - \frac{nl}{c}), \quad (6)$$

where $\delta\hat{E}_{THz,\text{in}}(t) \equiv \delta\hat{E}_{THz}(z=-l/2,t)$ and

$$f(t) = \frac{dl}{nc}\frac{\partial}{\partial t}E_{THz}(z=l/2,t) \quad (7)$$

recovers the second part of Eq. (1) in the main text. Calculating the root-mean-square (rms) standard deviation $\Delta E_{rms}(t) \equiv \langle\delta\hat{E}^2_{THz,\text{out}}\rangle^{1/2}(t)$ at the end of the nonlinear section results in

$$\Delta E_{rms}(t) = e^{f(t)} \sqrt{\left\langle \delta \hat{E}^2_{THz,\text{in}}(t - \frac{nl}{c}) \right\rangle}. \qquad (8)$$

The first part of Eq. (1) in the main text follows from a bare vacuum or fully coherent input, as in our experiment. In this case, $\langle \delta \hat{E}^2_{THz,\text{in}} \rangle^{1/2}(t)$ is given by the rms vacuum electric field $\Delta E_{vac}$.

**Time-domain noise patterns and temporal changes of the local phase velocity:**

The Pockels effect[33] causes a change in refractive index $\Delta n = rn^3 \cdot E_{THz}$ with the effective electro-optic coefficient $r = -d/n^4$ linking $d \cdot \partial E_{THz}/\partial t$ and therefore $f(t)$ to accelerations and retardations of the local reference frame. The linear refractive index is defined as the ratio between the velocity of light in vacuum $c$ and the local phase velocity $v_{loc}$, i.e. $n(t) = c/v_{loc}(t)$. Together with Eq. (1), we find

$$f(t) = -\frac{l}{c}\frac{\partial n}{\partial t} = l\frac{n^2}{c^2}\frac{\partial v_{loc}}{\partial t}. \qquad (9)$$

This expression is of general character as it does not depend on the specific nonlinearity that is used to induce the phase shifts which result in squeezing of the electromagnetic field and ultimately the emission of nonclassical radiation. For example, analogous noise patterns as the ones found in our experiments might result from direct modulation of $v_{loc}$ by the near-infrared pump intensity $I_p(t)$ via third-order effects causing a nonlinear index of refraction $n_2$ and therefore $\Delta n(t) = n_2 \cdot I_p(t)$. Definitely, it is clear from Eqs. (1) and (9) that excess noise with respect to the bare vacuum level may be traced to acceleration of the local reference frame, i.e. $\partial v_{loc}/\partial t > 0$. On the other hand, retardation with $\partial v_{loc}/\partial t < 0$ underlies a decrease of the local quantum fluctuations. These facts lead us to suggest a generalized understanding of quadratures in a time-domain context, as outlined in the conclusion.